\documentclass[useAMS,usenatbib]{mn2e}

\usepackage{times}
\usepackage{graphicx}

\def\gsim{\mathrel{\raise0.35ex\hbox{$\scriptstyle >$}\kern-0.6em
\lower0.40ex\hbox{{$\scriptstyle \sim$}}}}
\def\lsim{\mathrel{\raise0.35ex\hbox{$\scriptstyle <$}\kern-0.6em
\lower0.40ex\hbox{{$\scriptstyle \sim$}}}}
\def\m@th{\mathsurround=0pt }
\def\eqalign#1{\null\,\vcenter{\openup1\jot \m@th
 \ialign{\strut\hfil$\displaystyle{##}$&$\displaystyle{{}##}$\hfil
 \crcr#1\crcr}}\,}

\newcommand{\AIPS}{{$\cal AIPS\/$}}

\title[VLBI imaging of SMGs]
      {Deep, ultra-high-resolution radio imaging of submillimetre galaxies
       using Very Long Baseline Interferometry}
\author[Biggs, Younger \& Ivison]
       {A.\,D.\ Biggs,$\!^1$\thanks{E-mail: abiggs@eso.org}
        J.\,D.\ Younger$^2$\thanks{Hubble Fellow} and
        R.\,J.\ Ivison$^{3,4}$
\vspace*{1mm}\\
$^1$ European Southern Observatory, Karl-Schwarzschild-Strasse 2,
     D-85748 Garching, Germany\\
$^2$ School of Natural Sciences, Institute for Advanced Study, Einstein Drive, Princeton NJ 08540, USA \\
$^3$ UK Astronomy Technology Centre, Royal Observatory, Blackford Hill,
     Edinburgh EH9 3HJ\\
$^4$ Institute for Astronomy, University of Edinburgh, Royal Observatory,
     Blackford Hill, Edinburgh EH9 3HJ}

\date{\fbox{\sc Draft dated: \today\ }}

\pagerange{000--000}

\begin{document}

\maketitle

\begin{abstract}
We present continent-scale very-long-baseline interferometry (VLBI) --
obtained with the European VLBI Network (EVN) at a wavelength of 18~cm
-- of six distant, luminous submillimetre-selected galaxies
(SMGs). Our images have a synthesized beam width of
$\approx$30\,milliarcsec {\sc fwhm} -- three orders of magnitude
smaller in area than the highest resolution Very Large Array (VLA)
imaging at this frequency -- and are capable of separating radio
emission from ultra-compact radio cores (associated with active
super-massive black holes -- SMBHs) from that due to starburst
activity. Despite targeting compact sources -- as judged by earlier
observations with the VLA and MERLIN -- we identify ultra-compact
cores in only two of our targets. This suggests that the radio
emission from SMGs is produced primarily on larger scales than those
probed by the EVN, and therefore is generated by star formation rather
than an AGN -- a result consistent with other methods used to identify
the presence of SMBHs in these systems.
\end{abstract}

\begin{keywords}
   galaxies: starburst -- galaxies: formation -- cosmology:
   observations -- cosmology: early Universe
\end{keywords}

\section{Introduction}
\label{sec:intro}

With bolometric luminosities rivalling quasars, submillimetre galaxies
(SMGs) are some of the most extreme objects in the Universe \citep[for
a review, see][]{blain2002}, yet they outnumber similarly luminous
quasars by many orders of magnitude \citep{chapman2005}. Selected via
their dust-reprocessed thermal emission, they are thought to be
powered by intense bursts of star formation triggered by major mergers
of gas-rich galaxies \citep{tacconi2006, tacconi2008,
younger2008highres,younger2010highres} at $z\sim 2-3$ \citep{chapman2005} with a
significant tail extending out to higher redshifts \citep{eales2003,
younger2007, younger2009.aztecsma, wang2007, wang2008, greve2008,
capak2008, schinnerer2008, daddi2009, coppin2009}, and are
the likely progenitors of massive galaxies in the local Universe
\citep{scott2002, blain2004, smail2004, swinbank2006}. In addition,
it is thought that infrared- (IR-)luminous objects ($L_{\rm IR} \gsim
10^{12-13} \mathrm{L}_{\sun}$) come to dominate the cosmic star-formation rate
density at high redshift \citep[$z\sim 2-4$;][]{blain1999,
hopkins2009.ulirg, hopkins2009.ulirg2}, making the SMG population an
important contributor to the build-up of stellar mass during the epoch
of galaxy formation.

At the same time, over the past twenty years it has become clear that
super-massive black holes (SMBHs) in the nuclei of galaxies are
common, even at high redshift \citep[e.g.][]{ivison2008} and that
their masses are strongly correlated with the stellar component of
their host galaxies \citep[e.g.,][]{kormendy1995, magorrian1998,
  Gebhardt2000, Ferrarese2000, Tremaine2002, novak2006,
  Hopkins2007obs}. These correlations are indicative of a close and
fundamental link between SMBHs and stellar populations, and the
enormous difference ($\sim 10^9$) in linear scales suggests that this
may be accomplished via accretion-related radio jets or
radiatively-driven winds.  Recent theoretical models cast the
co-evolution of SMBHs and galaxies in the context of a cosmic cycle
driven by major mergers \citep[][and references within]{hopkins2007a,
  hopkins2007b}, wherein gas-rich, spiral galaxies collide and
trigger a gas inflow, thereby fuelling a nuclear starburst
\citep{hernquist1989a, mihos1994, mihos1996} and feedback-regulated
SMBH growth \citep{silk1998, Page2004, diMatteo2005,
  hopkins2007theory, younger2008.smbh}, eventually revealing a bright
quasar, after which gas exhaustion and violent relaxation transform
the remnant into a red elliptical galaxy \citep{barnes1992b}.

These models of galaxy evolution, which match constraints from both
galaxy \citep{hopkins2006} and SMBH populations \citep{robertson2006a,
hopkins2007theory, younger2008.smbh}, predict that hyperluminous
starbursts at high redshift will be associated with periods of active
SMBH growth. SMGs are thus likely candidates for the transition
objects that theory predicts should be powered by a mixture of star
formation and active black hole growth \citep[see
also][]{younger2009.warmcold}.  To date, searches for actively growing
SMBHs in SMGs have focused on X-ray observations \citep{alexander2005,
alexander2008} and mid-IR photometry and spectroscopy
\citep{ivison2004, lutz2005, menendez2007, menendez2009, valiante2007,
pope2008b}. Whilst promising, these approaches are often compromised:
even hard X-ray searches will miss Compton-thick SMBHs, and the
interpretation of mid-IR spectra can be very sensitive to modelling
assumptions, especially regarding the origin of power-law spectra
\citep{yun2001, younger2009.warmcold}.

Very-high-resolution radio imaging, provided by continent-scale
very-long-baseline interferometry (VLBI), avoids many of these issues;
even in the most extreme environments, the interstellar medium is
optically thin to radio emission and the dominant radio emission
mechanisms are well understood. In fact, there is a well-known upper
limit to the brightness temperature ($T_{\rm b}$) a starburst can
achieve \citep[$T_{\rm b} \lsim 10^5$ {\sc k} --][]{condon1991b}, and
any compact object detected at VLBI resolution will exceed (or be
close to) this limit.

A number of authors have used this technique in order to distinguish
between AGN and starburst activity in both high-redshift QSOs
\citep{beelen2004,momjian2005,momjian2007} and in ultra-luminous
infrared galaxies (ULIRGs), the low-redshift analogues of the
SMGs. The latter are particularly important as they present a far more
detailed view of the relationship between the star formation and AGN
emission in an intense starburst than will ever be possible in a
distant SMG, and aid in the interpretation of these less spatially
resolved sources; the synthesised beam of a typical VLBI array is high
enough, for example, to resolve individual supernova remnants in local
ULIRGs e.g.\ Arp~220 \citep{lonsdale2006} and IRAS~17208-0014
\cite{momjian2006}. Examples of ULIRGs that are believed to contain an
AGN include Arp~220 \citep{downes2007} and Mrk~273 \citep{carilli2000}.

Only one VLBI observation of a classical, high-redshift,
optically-faint SMG \citep{momjian2010} has been published to date,
continuum emission from GOODS~850-3 having been detected in a tapered
High Sensitivity Array (HSA) image. In this paper we present the first
VLBI survey of SMGs, using the European VLBI Network (EVN) to search
for ultra-compact cores in a sample of objects in the Lockman
Hole. These data, with a synthesized beam size of
$\approx$30\,milliarcsec (mas) {\sc fwhm}, are the highest-resolution
radio detections of SMGs ever achieved and have allowed us to put the
tightest constraints to date on the brightness temperatures of a
significant sample of high-redshift starbursts.

This paper is organised as follows: in \S\ref{sec:targets} we describe
our target selection and in \S\ref{sec:reduce} give details of the
observing strategy, data reduction and imaging. In \S\ref{sec:results}
we present our high-resolution images of each SMG and describe each
source in detail. In \S\ref{sec:discuss} we discuss our findings
before presenting our conclusions in \S\ref{sec:conclude}.

Where necessary we have assumed a flat $\Lambda$CDM cosmology of
$\Omega_{\Lambda} = 0.73$, $\Omega_{m} = 0.27$ and $H_0 =
71$~km\,s$^{-1}$\,Mpc$^{-1}$ \citep{hinshaw2009}.

\section{Target Selection and Auxiliary Data}
\label{sec:targets}

\begin{table}
\begin{center}
\caption{Names of SMGs targeted by our EVN observations. The six are
  split into two subsamples, the first three being significantly
  brighter.}
\begin{tabular}{ccccc} \hline
Name  & SHADES     & 8-mJy        & BOLOCAM      & MAMBO       \\ \hline
SMG06 & LOCK850.30 & LE850.12     & \ldots       & \ldots      \\
SMG10 & \dots      & \ldots       & \ldots       & LH1200.008  \\
SMG11 & \ldots     & \ldots       & 1100.003a    & \ldots      \\ \hline
SMG01 & LOCK850.01 & LE850.01     & 1100.014     & LH~1200.005 \\
SMG02 & LOCK850.04 & LE850.14     & \ldots       & LH~1200.003 \\
SMG04 & LOCK850.16 & LE850.07     & \ldots       & LH~1200.096 \\ \hline
\end{tabular}
\label{tab:targetnames}
\end{center}
\end{table}

\begin{figure*}
\begin{center}
\includegraphics[trim= 10mm 10mm 0mm 0mm, clip, scale=1.0]{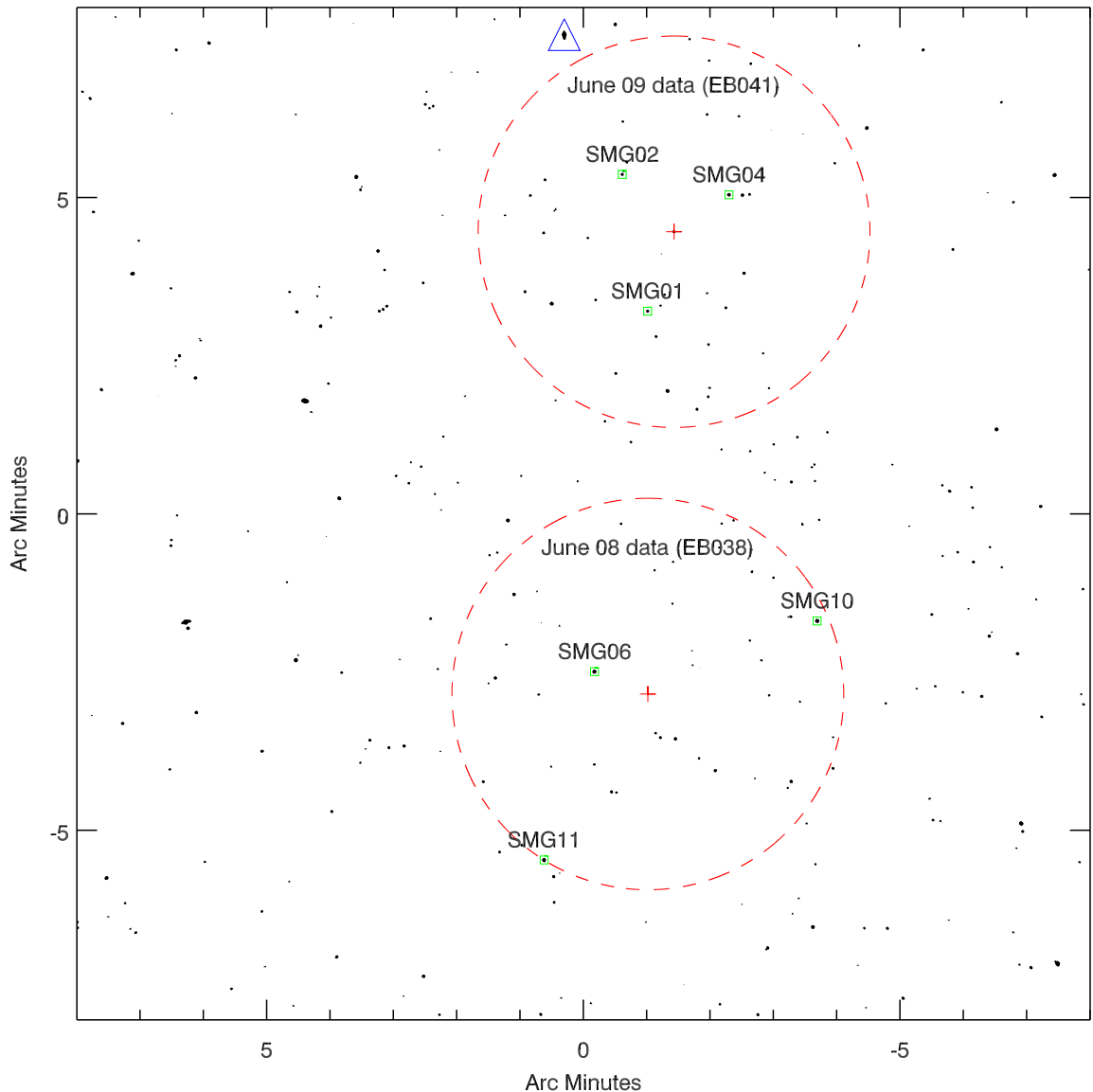}
\end{center}
\caption{VLA 1.4-GHz map of the Lockman Hole showing the locations of
  the six targeted SMGs and their division into two separate observing
  epochs. The dashed circles correspond to the primary beam of the
  Effelsberg telescope at 18\,cm centred on the pointing positions of
  all telescopes apart from the Westerbork Synthesis Radio Telescope
  (WSRT) which pointed at each SMG consecutively; the small green
  boxes around each SMG position are approximately the size of the
  WSRT field of view (11\,arcsec). The pointing positions (red
  crosses) for all telescopes (apart from WSRT) were
  10$^{\mathrm{h}}$\,52$^{\mathrm{m}}$\,01\fs250,
  57$^{\circ}$\,18$^{\prime}$\,42\farcs76 for EB038 and
  10$^{\mathrm{h}}$\,51$^{\mathrm{m}}$\,58\fs150,
  57$^{\circ}$\,26$^{\prime}$\,01\farcs13 for EB041. The triangle just
  outside the EB041 field marks the source that was detected using the
  SMG02 data.}
\label{fig:field}
\end{figure*}

\begin{table*}
\begin{center}
\caption{Observational data of our sample. 850-$\mu$m data are
  taken from \citet{coppin2006}, 1100-$\mu$m data from
  \citet{laurent2005} and those at 1200~$\mu$m from 
  \citet{greve2004}. 350-$\mu$m flux densities are taken from
  \citet{laurent2006}. The positions were measured from VLA 1.4-GHz
  imaging \citep{ivison2002, biggs2006} and all redshifts are
  spectroscopic. The VLA total flux densities and the MERLIN peak flux
  densities have both been measured at 1.4\,GHz and 
  corrected for bandwidth smearing and primary beam attenuation.
  Due to the different antenna sizes, for the MERLIN data the attenuation
  has been calculated using a sensitivity-weighted average of each baseline.
  The MERLIN maps can
  display multiple components: the quoted fluxes refer to the
  brightest. The final column gives the spectral index between 1.4\,GHz
  (VLA) and 610\,MHz \citep[from imaging with the Giant Metre-wave Radio
  Telescope --][]{Ibar2009, Ibar2010} where $S_{\nu} \propto \nu^{\alpha}$.}
\begin{tabular}{lccccccccc} \hline
Name & $\alpha_{\mathrm{J2000}}$ & $\delta_{\mathrm{J2000}}$ & $S_{\rm 350 \mu m}$ & $S_{\rm 850\mu m}$ & $S_{\rm 1100|1200\mu m}$ & $z$ & $S_{1.4,\mathrm{VLA}}$ & $S_{1.4,\mathrm{MERLIN}}$ & $\alpha_{0.61}^{1.4}$\\
& ($^{\mathrm{h\,m\,s}}$) & ($^{\circ\,\prime\,\prime\prime}$) & (mJy) & (mJy) & (mJy) & & ($\mu$Jy) & ($\mu$Jy\,beam$^{-1}$) \\ \hline
SMG06 & 10:52:07.49 & +57:19:04.0 & $38.0\pm 7.2$ & $4.7^{+1.5}_{-1.6} $& \ldots & $2.689\pm0.01$ & $246\pm8$ & $296\pm8$ & $+0.83$ \\
SMG10 & 10:51:41.43 & +57:19:51.9 & \ldots & \ldots & $4.1\pm0.9$ & $1.212\pm0.01$ & $295\pm9$ & $157\pm10$ & $-0.69$ \\
SMG11 & 10:52:13.38 & +57:16:05.4 & \ldots & \ldots & $6.0\pm 1.4$ & \ldots & $245\pm5$ & $209\pm11$ & $-0.66$\\ \hline
SMG01 & 10:52:01.25 & +57:24:45.8 & $24.1\pm 5.5$ & $8.8\pm 1.0$ & $3.6\pm 0.6$, $4.4\pm 1.3$ & $3.38\pm0.02$ & $73\pm5$ & $60\pm9$ & $+0.21$ \\
SMG02 & 10:52:04.23 & +57:26:55.5 & $24.9\pm 9.1$ & $10.6^{+1.7}_{-1.8}$ & $3.6\pm 0.6$ & $1.482\pm0.01$ & $66\pm5$ & $64\pm10$ & $-0.56$ \\
SMG04 & 10:51:51.69 & +57:26:36.1 & $25.7 \pm 15.8$ & $5.8^{+1.8}_{-1.9}$ & $1.6\pm 1.6$ & $1.147\pm0.01$ & $110\pm9$ & $76\pm11$ & $-0.43$ \\ \hline
\end{tabular}
\label{tab:targetdata}
\end{center}
\end{table*}

The six targets were drawn from the \citet{biggs2008} sample of SMGs
in the Lockman Hole with high-resolution (synthesized beam size
$\approx$200--500\,mas) radio imaging provided by the Multi-Element
Radio-Linked Interferometer Network (MERLIN), an array of seven
telescopes across the United Kingdom. Table~\ref{tab:targetnames}
gives the various names of each SMG (most are detected in several
submm surveys at different wavelengths) whilst in
Table~\ref{tab:targetdata} we reproduce various observational data,
including the position of the radio counterpart, their submm and
1.4-GHz flux densities (measured both by MERLIN -- \citealt{biggs2008}
-- and the VLA -- \citealt{ivison2002, biggs2006}) and radio spectral
index \citep{Ibar2009, Ibar2010}. A spectroscopic redshift is given
where available. The first three SMGs were selected on the basis of
their high peak flux density in the MERLIN map, as well as the
presence of compact emission that could not be resolved at full MERLIN
resolution. In addition, with a maximum separation of 5.7\,arcmin,
they could be observed with a single pointing of the EVN (the primary
beam of the Effelsberg telescope is 6\,arcmin at 18\,cm; see
Fig.~\ref{fig:field}). In order to widen the range of luminosities in
the sample -- and hence to probe galaxies representative of the wider
SMG population -- the second group of SMGs are considerably fainter
(by 3$\times$, based on their VLA flux densities) and also detected by
MERLIN. The maximum separation between these three sources is only
2.2\,arcmin, so the primary beam attenuation is reduced compared to
EB038 -- useful when trying to detect such faint sources.

The 850-$\mu$m data for this sample was provided by the Submillimeter
Common User Bolometer \citep[SCUBA --][]{holland1999, coppin2006} on
the 15-m James Clerk Maxwell Telescope, the 350-$\mu$m data are from
SHARC-{\sc ii} imaging at the 10-m Caltech Submillimeter Observatory
\citep[CSO --][]{kovacs2006, coppin2008.sharcii}, and the millimetre
($\lambda \approx 1100-1200\mu$m) survey data \citep{greve2004,
laurent2005, laurent2006} are from the MAMBO bolometer
\citep{Kreysa1998} on the Institut de Radio Astronomie
Millim\'{e}trique's (IRAM's) 30-m telescope and from Bolocam
\citep{Haig2004} on the CSO.

\section{Observations, Data Reduction and Imaging}
\label{sec:reduce}

Our EVN observations of the Lockman Hole were split into two separate
observing sessions, the first targeting the brightest three SMGs
(project code EB038; 2008 June 3 and 5) and the second (EB041; 2009
June 5 and 6) targeting the remaining three
(Fig.~\ref{fig:field}). The EB038 and EB041 observations had durations
of 12 and 14\,hr. The seven telescopes which participated in EB038 are
detailed in Table~\ref{tab:telescopes} and include the 76-m Lovell
telescope at Jodrell Bank, the 100-m telescope at Effelsberg and the
Westerbork array of 14 $\times$ 25-m antennas. EB041 also included the
25-m Urumqi dish in Western China, this providing much longer
baselines and therefore tighter constraints on the brightness
temperature, $T_{\rm b}$. Due to a mechanical failure of the Lovell
telescope shortly after the beginning of EB041, this was replaced by
the 25-m Mark~{\sc ii} telescope (also located at Jodrell Bank) for
the vast majority of the project.

\begin{table}
\begin{center}
\caption{Details of the telescopes used in the EVN observing sessions
  in 2008 (EB038) and 2009 (EB041). The Lovell telescope suffered a
  mechanical failure shortly after beginning EB041 and only
  contributed about an hour of data before being replaced by the
  smaller Mark~{\sc ii} telescope.}
\begin{tabular}{ccc} \hline
Telescope & Diameter (m) & Project \\ \hline
Lovell, UK & 76 & EB038 \\
Mark II, UK & 25 & EB041 \\
Effelsberg, Germany & 100 & both \\
Westerbork, The Netherlands & $14 \times 25$ & both \\
Onsala, Sweden & 25 & both \\
Medicina, Italy & 32 & both \\
Noto, Italy & 32 & both \\
Torun, Poland & 32 & both \\
Urumqi, China & 25 & EB041 \\ \hline
\end{tabular}
\label{tab:telescopes}
\end{center}
\end{table}

The faintness of the targets made phase referencing essential and for
our calibrator we used the unresolved $\sim$200-mJy source,
J1058+5628, which is located only one degree from the SMGs. For EB038
we used a cycling time of 300\,s, but increased this to 600\,s for
EB041. As explained in \S\ref{sec:targets}, the targets lie within the
primary beams of the largest telescope and so could be observed in a
single pointing. An exception was the WSRT which in phased-array mode
has a field of view of only 11\,arcsec. Therefore, this telescope was
cycled consecutively through the three SMG positions, changing after
every second SMG scan. This strategy required that the data be
correlated three times, once at each of the three WSRT pointings. The
bright quasars 3C\,345 and 4C\,39.25 were both observed for 600\,s as
bandpass and delay calibrators.

Both EB038 and EB041 were observed with dual circularly-polarised
receivers tuned to a wavelength close to 18\,cm. The total bandwidth
was 64~MHz (8$\times$8-MHz bands) sampled at 2~bits for a total data
rate of 512 Mbits\,s$^{-1}$. The data were correlated at the Joint
Institute for VLBI in Europe (JIVE) with the EVN Mark {\sc iv} Data
Processor; bandwidth and time smearing losses were restricted to
$\sim$10~per~cent at a radius of $\approx$1\,arcmin from the phase
centre of each image by averaging the correlator output every 1\,s and
splitting each 8-MHz band into 64 channels. Unfortunately, during the
correlation of data for SMG02, the disk containing the data from the
Effelsberg telescope failed. As a result, only one quarter of its data
were salvaged and the sensitivity of the images of this source and of
SMG04 are significantly worse than those of the other targets.

The data for each target/epoch were reduced using \AIPS\ in the
following way: initial amplitude calibration was accomplished by
applying system temperatures measured during the observations, with
the standard gain curves of each antenna. After finding and removing
corrupted data (as well as the WSRT scans corresponding to the other
two SMG positions) the delay error on each antenna was calculated from
a small portion of 4C\,39.25 data and removed from the entire
dataset. The same source was also used to measure and flatten the
bandpass shape in both amplitude and phase. Residual delay, rate and
phase solutions were then calculated for the phase calibrator and
interpolated onto the target SMG. Finally, time-dependent amplitude
gain corrections were determined for J1058+5628 and applied to the
target data.

Once calibrated, the target data were imaged using {\sc imagr} and
naturally weighted to ensure maximum sensitivity. For the EB038 data,
the combination of this and the $uv$ coverage resulted in a nearly
circular synthesised beam of $27 \times 25$~mas$^2$ and an
r.m.s.\ noise level of 10.5\,$\mu$Jy\,beam$^{-1}$. For the EB041 data,
we found that applying a Gaussian taper to the weights (with a value
of 30~per~cent at a distance of 5~M$\lambda$ in both the $u$ and $v$
coordinates) produced the best images. This produces r.m.s.\ noise
levels between 10.7 and 14.3~$\mu$Jy\,beam$^{-1}$ and a synthesised
beam with dimensions of $23 \times 22$~mas$^2$.

\begin{figure*}
\begin{center}
\includegraphics[scale=0.293]{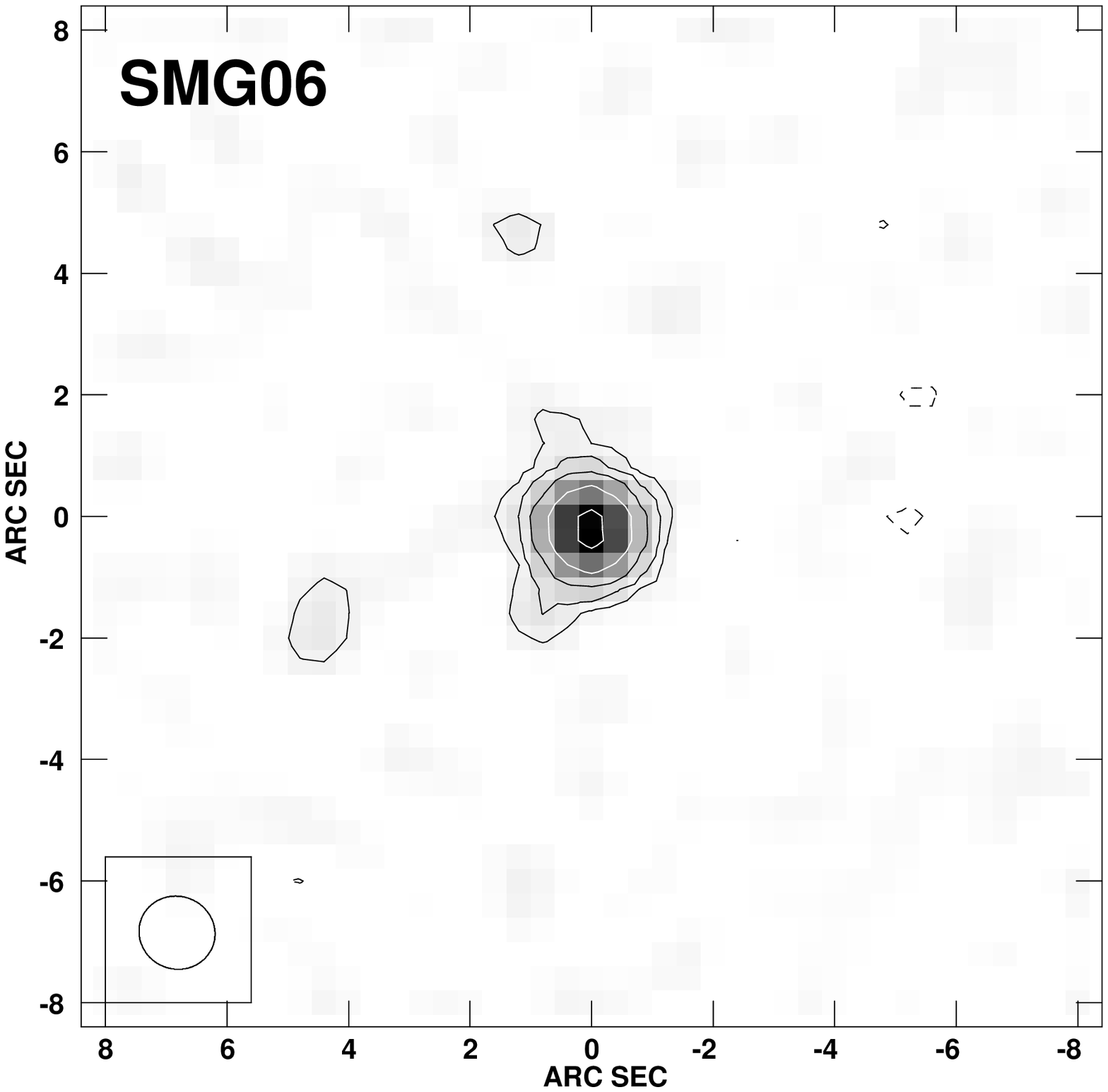}%
\includegraphics[scale=0.3]{SMG2_MER.PS}%
\includegraphics[scale=0.3]{SMG2_EVN.PS}
\includegraphics[scale=0.293]{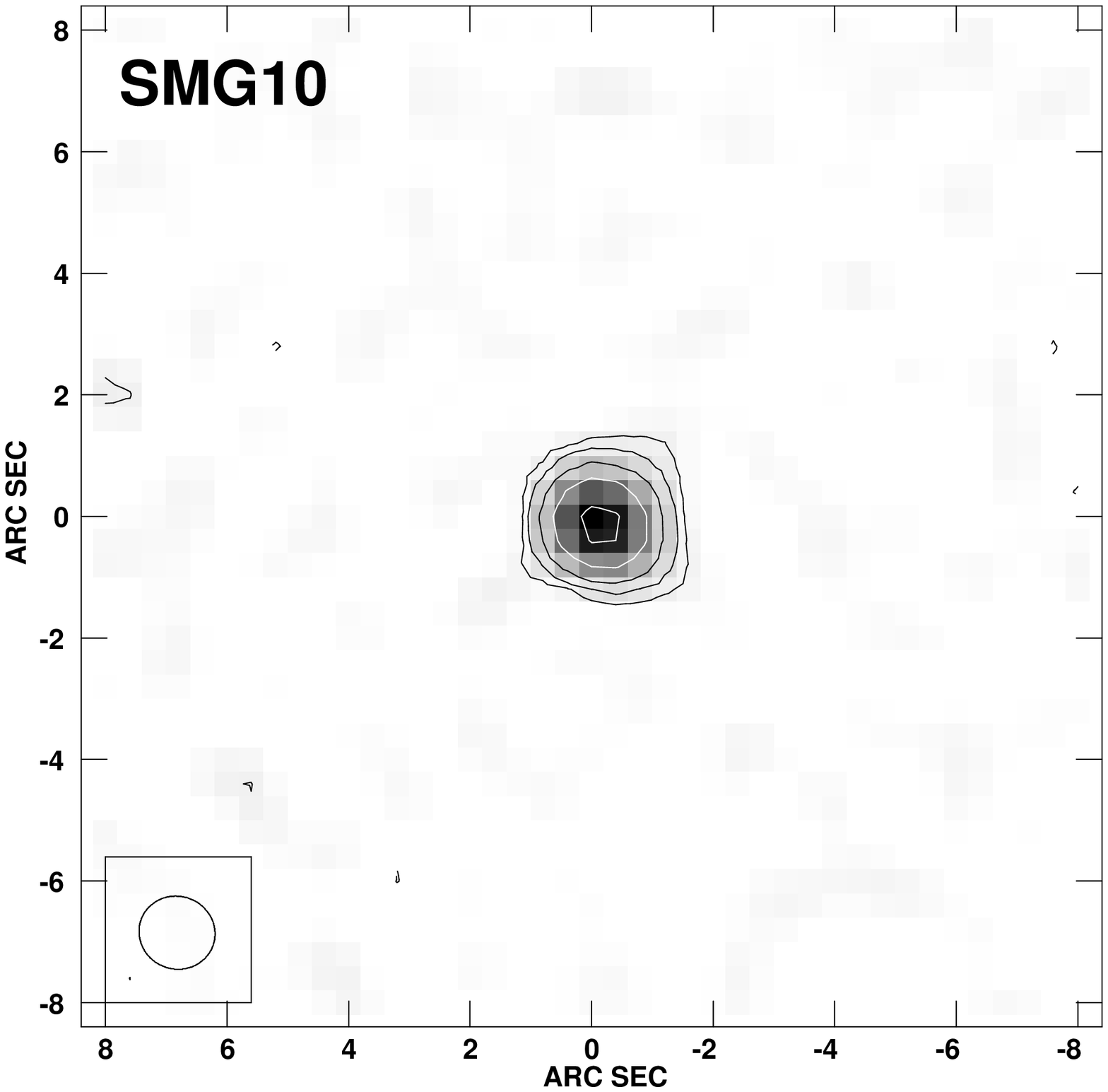}%
\includegraphics[scale=0.3]{SMG1_MER.PS}%
\includegraphics[scale=0.3]{SMG1_EVN.PS}
\includegraphics[scale=0.293]{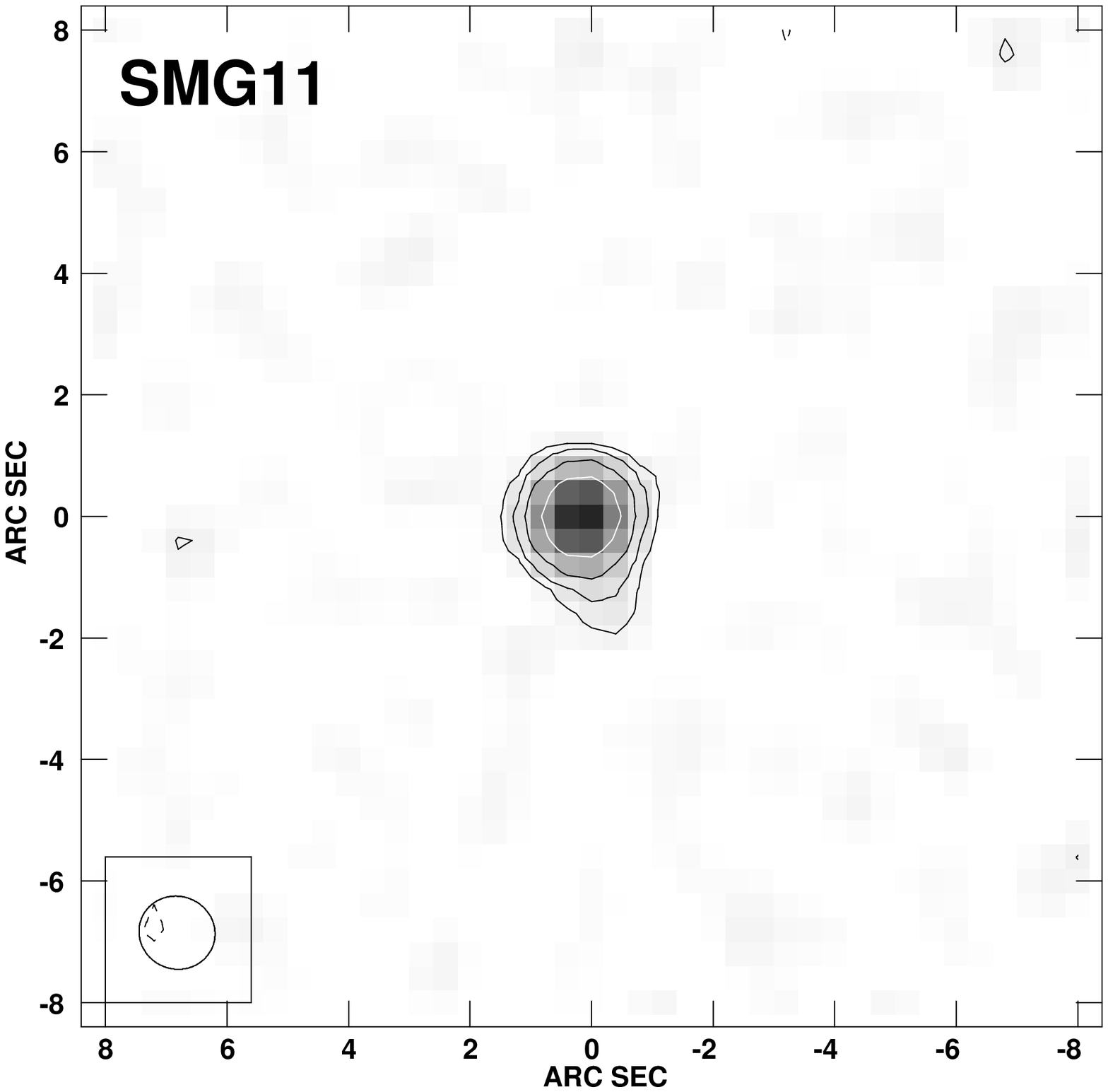}%
\includegraphics[scale=0.3]{SMG3_MER.PS}%
\includegraphics[scale=0.3]{SMG3_EVN.PS}
\caption{Images of our sample of SMGs. {\sc Left column:}
  robustly-weighted VLA images at 21~cm, {\sc centre column:}
  naturally-weighted MERLIN images at 21~cm, {\sc right column:}
  naturally-weighted EVN images at 18~cm. For the VLA maps, contours
  are plotted at $-3, 3, 6, 12, 24, 48\times$ the r.m.s.\ noise. The
  MERLIN and EVN maps have contours at $-3, 3, 4, 5, 10, 20, 30,
  40\times$ the r.m.s.\ noise (Table~\ref{tab:evn}). Synthesised beams
  are shown in the bottom-left corners and each image is the same
  multiple of the beam area. Note that the MERLIN and EVN images shown
  here have not been corrected for primary-beam attenuation.}
\label{fig:images}
\end{center}
\end{figure*}

\begin{figure*}
\begin{center}
\includegraphics[scale=0.293]{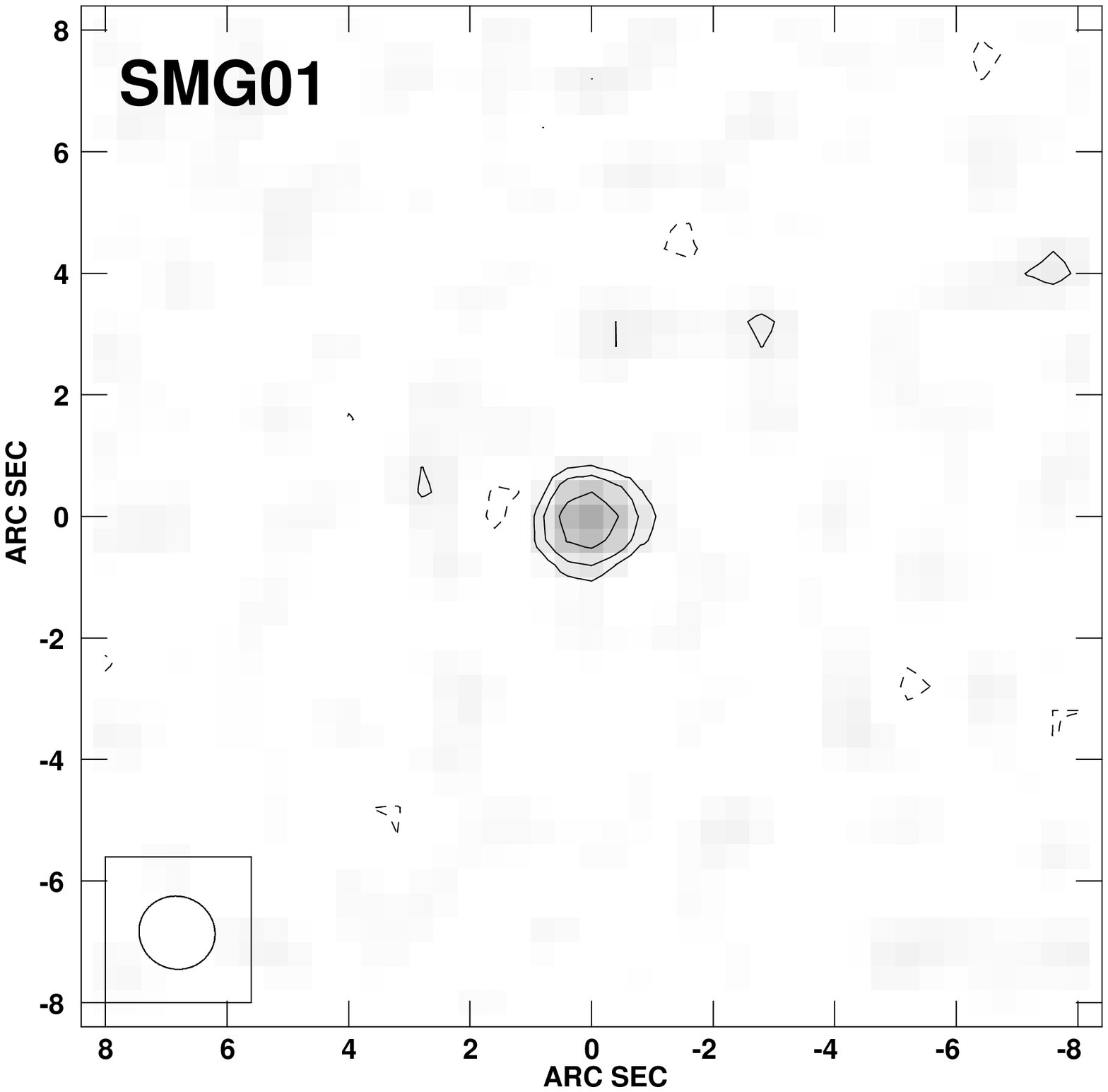}%
\includegraphics[scale=0.3]{SMG4_MER.PS}%
\includegraphics[scale=0.3]{SMG4_EVN.PS}
\includegraphics[scale=0.293]{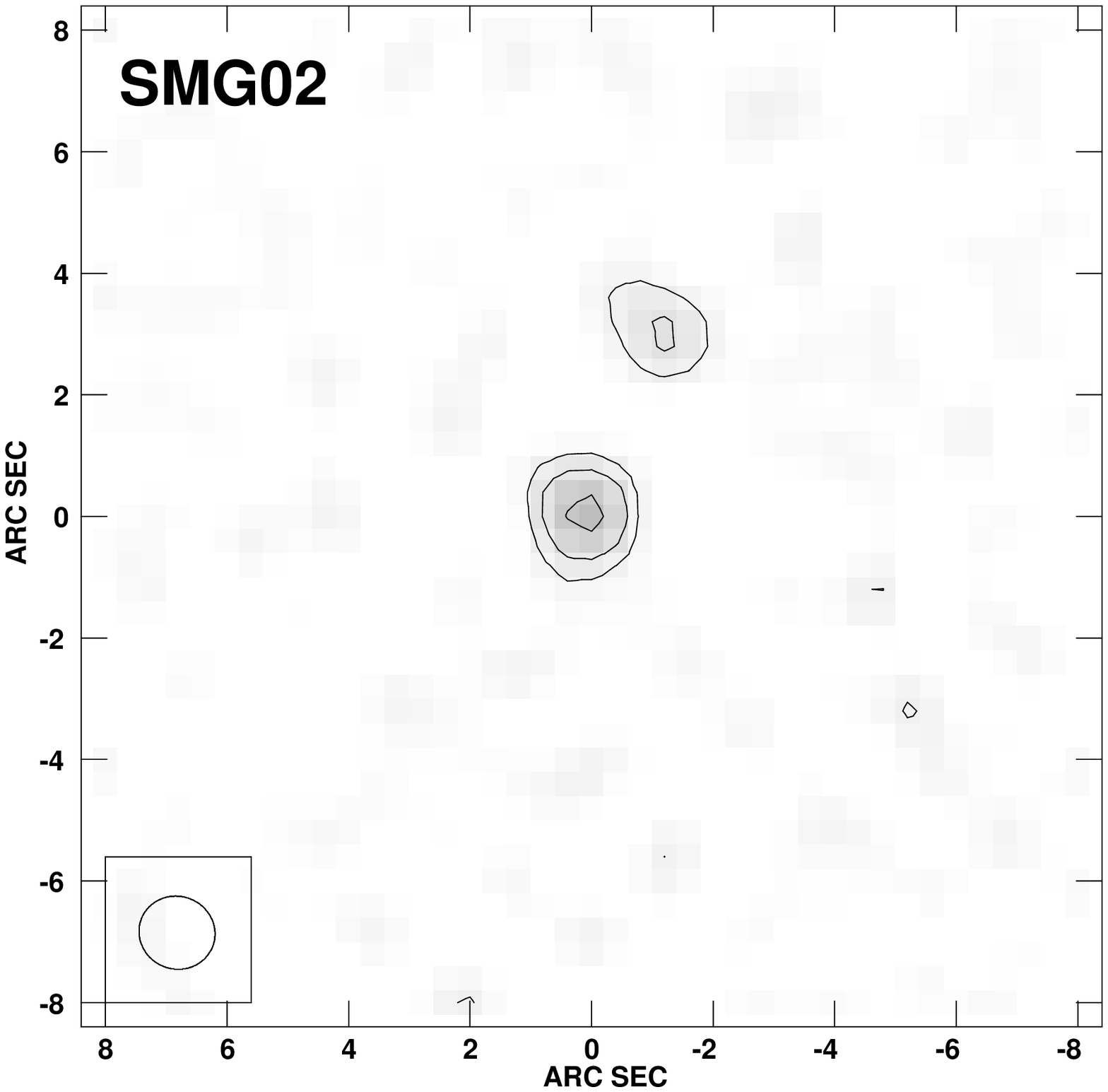}%
\includegraphics[scale=0.3]{SMG5_MER.PS}%
\includegraphics[scale=0.3]{SMG5_EVN.PS}
\includegraphics[scale=0.293]{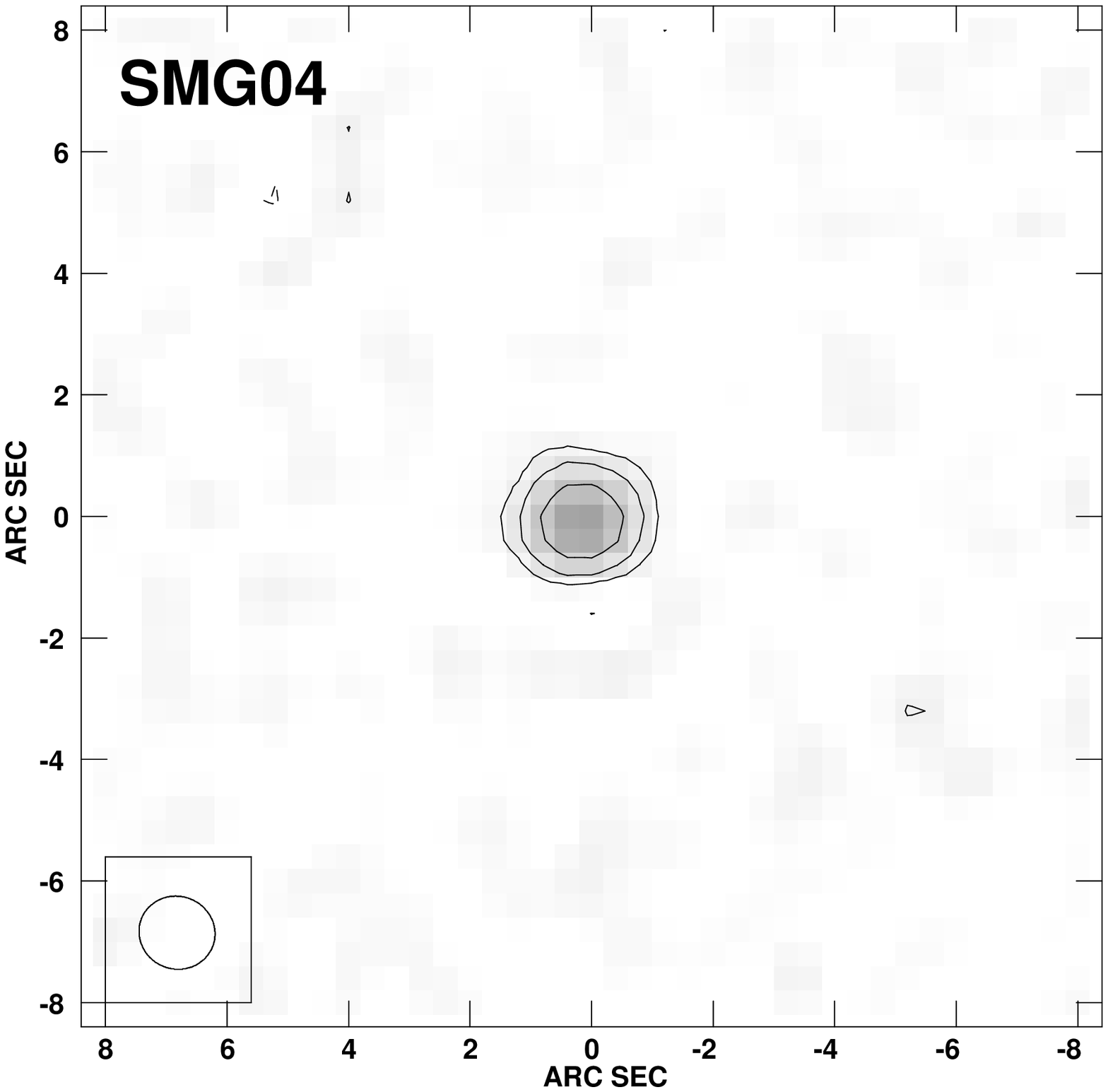}%
\includegraphics[scale=0.3]{SMG6_MER.PS}%
\includegraphics[scale=0.3]{SMG6_EVN.PS}
\contcaption{}
\end{center}
\end{figure*}


\section{Results}
\label{sec:results}

The final images are shown in Fig.~\ref{fig:images}. Of the six SMGs
that we targeted, two (SMG06 and SMG11) have been detected. Of the
other four, we have searched for emission on larger scales by tapering
the data in the ($u,v$) plane. The shortest baseline in our EVN array
(between the WSRT and Effelsberg) is also the most sensitive and
corresponds to emission on scales of approximately 100~mas, still a
factor of four smaller in beam area than that of the MERLIN
maps. This, combined with the tapered maps having sensitivities that
are poorer by about 50~per~cent and the higher sidelobes in the
synthesised beam, resulted in the continued non-detection of all SMGs
other than SMG06 and 11.

\begin{figure*}
\begin{center}
\includegraphics[scale=0.3]{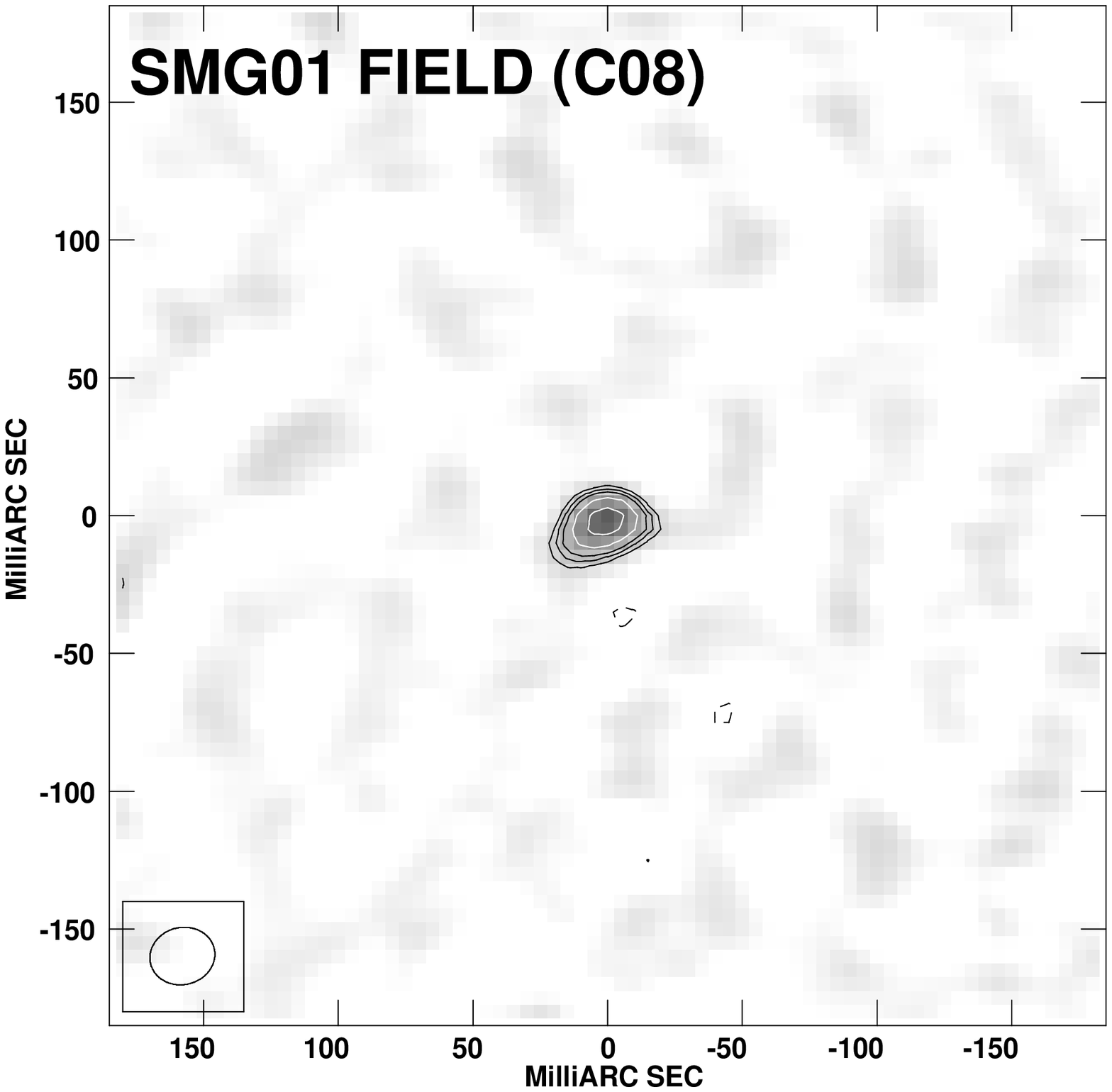}
\includegraphics[scale=0.3]{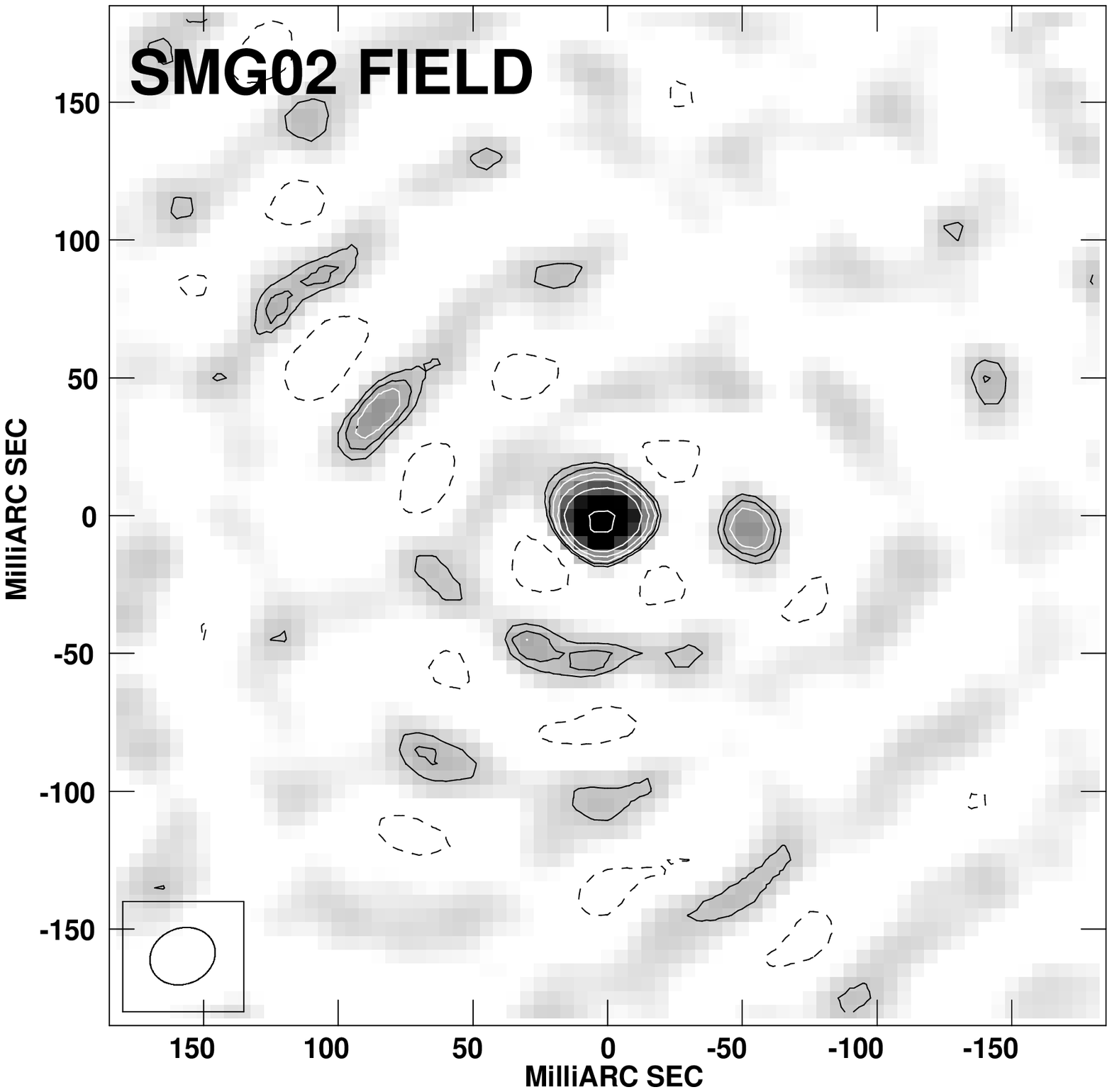}
\includegraphics[scale=0.3]{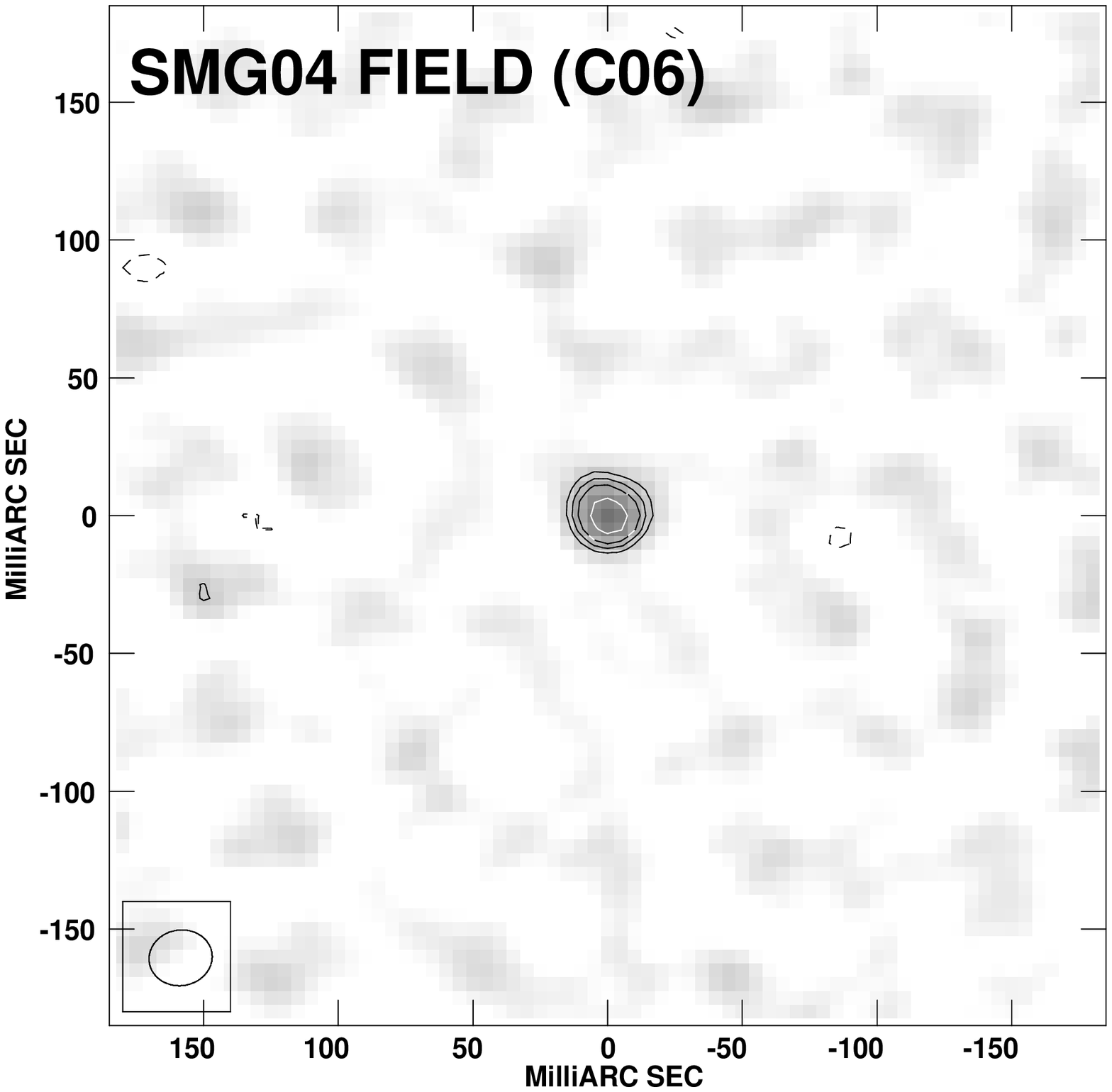}
\caption{EVN 18-cm maps of VLA- and MERLIN-detected sources that lay
  within the primary beam of the EB041 observations. Two of the
  sources are included in the control sample of \citet{biggs2008} and
  are labelled as such in the figures -- C06 was imaged from the SMG01
  data and C08 from the SMG04 data. The source in the SMG02 field was
  not included in the \citet{biggs2008} sample, but is marked as a
  triangle near the top of Fig.~\ref{fig:field}. These detections
  demonstrate that our observational and data reduction procedure is
  sound i.e.\ the lack of detections of SMGs from the faint sample is
  due to their faintness. The synthesised beams in these images have
  dimensions of approximately $24 \times 20$~mas$^2$ and contours are
  plotted at $-3, 3, 4, 5, 7, 10, 20\times$ the r.m.s.\ noise
  (16--20\,$\mu$Jy\,beam$^{-1}$). Data from the Urumqi telescope are not
  included so as to minimise the effects of smearing. The WSRT data
  are also not included as all sources are outside its primary beam.}
\label{fig:evnwide}
\end{center}
\end{figure*}

The lack of detections from the EB041 data might be ascribed to
problems with the observations (especially the phase referencing)
and/or the data reduction. However, as our data were correlated with a
large number of channels and short integration times we can search for
emission from sources elsewhere in each SMG's field. Three clear
detections were made, one in each field and are shown in
Fig.~\ref{fig:evnwide}. We can be confident therefore that the phase
referencing was successful and that the failure to detect emission
from SMG01, SMG02 and SMG04 is due to the lack of suitably-bright
compact emission in these sources.

For the two detections, we have measured their flux densities using
{\sc jmfit} in \AIPS. The fitted flux density values need to be
corrected for the attenuation of the primary beam and, as with the
MERLIN flux densities, this is not straightforward because of the
different diameters of the telescopes. We have therefore taken the
weighted mean of the attenuation factors for each baseline, assuming
that the primary beam of each is Gaussian in form and using the
\citet{strom2004} formulism for the primary beam of an interferometer
pair. The resulting values and the corrected peak flux densities are
both shown in Table~\ref{tab:evn}.

The attenuation factors are
generally insignificant, apart from SMG10 and SMG11. As already
noted, all MERLIN flux densities have been corrected for bandwidth
smearing and primary beam attenuation, the latter calculated in a
similar fashion to the EVN values. The pointing centre of the MERLIN
observations was the centre of Fig.~\ref{fig:field} and so the
attenuation factors are in general larger than for the EVN.

We now give brief descriptions of each source:

\noindent {\bf SMG01:} The first source of the faint subsample is
actually the brightest submm source in the Lockman Hole East region
(850.01). Recent {\it Spitzer} mid-IR spectroscopy has determined a
reliable redshift, $3.38\pm0.02$ \citep{coppin2010}. SMG01 is detected
as a single component by MERLIN with a 1.4-GHz flux of 90~$\mu$Jy, a
morphology that is consistent with its flat spectrum,
$\alpha=+0.21$. This would have corresponded to a 9-$\sigma$ detection
were it unresolved at the EVN resolution, but only weak (3-$\sigma$
i.e.\ $\sim$30\,$\mu$Jy\,beam$^{-1}$ peak) emission is seen near
the centre of the map.

\noindent {\bf SMG02:} The MERLIN map shows two distinct components,
but neither are detected in the EVN image.

\noindent {\bf SMG04:} The MERLIN map is dominated by a single
component which is elongated slightly, north-south. There is no
significant detection in the VLBI map, although we note that there is
weak (3-$\sigma$ i.e.\ $\sim$45\,$\mu$Jy\,beam$^{-1}$ peak) emission
in the EVN map close to the expected position of the source.

\noindent {\bf SMG06:} This source is detected robustly by the EVN, is
found to be unresolved by {\sc jmfit} and has a flux density that is
significantly in excess of the VLA value, presumably due to the source
variability first noted by \citet{ivison2002}. The MERLIN image is
slightly resolved, but has a peak flux density
(296\,$\mu$Jy\,beam$^{-1}$) that is similar to the EVN value. SMG06
has a highly-inverted radio spectrum, $\alpha = +0.83$, as also noted
by \citet{ivison2002}.

\noindent {\bf SMG10:} Despite being the brightest radio source in our
sample (VLA flux density = 295\,$\mu$Jy), SMG10 is not detected by our
EVN observations. An examination of the MERLIN map shows that this
source contains two components, although the brighter of the two
dominates, with a peak flux density of 157\,$\mu$Jy\,beam$^{-1}$. It
is possible that the weaker of the two components represents a jet,
but it could equally be one component of a late-stage merger. The
radio spectral index of the entire system, $\alpha=-0.69$, is
consistent with an optically-thin radio jet or a canonical starburst
\citep{Ibar2010}. As noted by \citet{ivison2007c}, matched
high-resolution imaging at multiple frequencies are required to
disentangle the various emission components in sources like SMG10.

\noindent {\bf SMG11:} This source is also detected by the EVN, but at
a considerably lower significance than SMG06, partly due to the
primary-beam attenuation. Even taking this into account, only
38~per~cent of the VLA flux is recovered. The MERLIN image is
dominated by a single (resolved) component with a peak flux density of
209\,$\mu$Jy\,beam$^{-1}$. The EVN detection is also marginally
resolved, approximately along the north-south axis. The deconvolved
size along this axis is 20~mas which is equivalent to a spatial extent
of 165~pc at the assumed redshift of 2.4 for this source.

\section{Discussion}
\label{sec:discuss}

The power of true, continent-scale VLBI instruments like the EVN to
identify actively growing SMBHs lies in their ability to distinguish
between different emission mechanisms via $T_{\rm b}$, separating
ultra-compact radio cores from more extended emission
\citep{Lonsdale1993}. \citet{condon1991b} showed that there exists a
maximum brightness temperature, $T_{\rm b}$, for radio emission
associated with a starburst. Assuming a canonical starburst
i.e.\ synchrotron, with a spectral index of $-0.8$, an electron
kinetic temperature of $10^4$~K and scaling the observed temperature
to the rest-frame of the SMG, we derive maximum values of $T_{\rm b}$
of approximately $5 \times 10^4$~K for all our sources. Both of our
EVN detections shown in Fig.~\ref{fig:images} have $T_{\rm b}\gsim
10^5$~K (Table~\ref{tab:evn}), significantly above this critical value
and both are therefore extremely likely to be produced by an active
SMBH.

For the SMGs that are not detected, Table~\ref{tab:evn} shows that at
most one half of the radio emission detected by the VLA could be
contained within a compact -- i.e.\ VLBI -- component and therefore that the
implied upper limits on $T_{\rm b}$ are rather low, $<10^5$~K for
each. Although the limits are not quite low enough to definitively
rule out the presence of a radio-emitting AGN in these SMGs, if such a
component is present it must be very near the level of detectability
in our EVN images.

What is clear though is that for these four systems the bulk of the
radio emission must arise on scales larger than the EVN beam. This can
be attributed to several mechanisms: cosmic rays accelerated by
supernova remnants created by short-lived, massive stars
\citep[i.e.\ star-forming galaxies obeying the far-IR/radio
  correlation:][]{condon1992, thompson2006, Ibar2008, lacki2009a,
  lacki2009b, ivison2010a}, or low-surface-brightness AGN-driven radio
emission, e.g.\ jets. Given that the presence of an AGN seems unlikely
given the non-detection of compact components in 4 out of 6 of the
SMGs, the evidence favours the first explanation, i.e.\ that the radio
emission in these objects is predominantly produced by a
starburst. Furthermore, since this sample was selected for compact emission at
$\sim$300-mas resolution and as most SMGs display extended radio
emission at that resolution \citep{biggs2008}, these results suggest
that the radio luminosity produced by SMGs is generally dominated by
star formation, a similar finding to that of \citet{momjian2010} who,
in the only other published VLBI observation of an SMG to date, find
that GOODS~850-3 is also consistent with star formation being the
origin of the radio emission.

It is also possible that the clearest case of an AGN detection in our
sample, SMG06, may not actually be a starburst galaxy at all, but a
highly-obscured AGN-dominated system. The presence of an AGN is
virtually guaranteed by the source variability that results in the EVN
flux being in excess of that detected by the VLA, but the large EVN
flux also leaves very little or no flux that might originate from a
powerful starburst. In this scenario, the slight resolution of the
MERLIN image might be due to a radio jet that is resolved out by the
EVN beam although we stress that the evidence for resolution in the
MERLIN image is slight, the total flux only exceeding the peak flux by
9~per~cent. However, given that the source has clearly brightened
since the VLA observations in 2001, simultaneous EVN and VLA imaging
might have detected a clear excess of flux with the larger beam.

\begin{table*}
\begin{center}
\caption{Measured noise levels, flux densities and brightness
  temperatures for our six targets. Where the target was not detected,
  we give a 3-$\sigma$ upper limit. Column 3 contains the primary beam
  attenuation which is used to correct the values given in the
  following columns. Brightness temperatures have been corrected for
  redshift ($T_{\rm b} = (1+z) T_{\rm b, obs}$) using the
  spectroscopic redshift except for SMG11 where we have assumed the
  median SMG redshift of $z = 2.4$ \citep{chapman2005}. These
  redshifts have also been used to calculate \citep{wright2006} the
  spatial extent of the beam at the distance of the source.}
\begin{tabular}{cccccccc} \hline
Name & r.m.s.\ noise & Primary beam & EVN peak flux & EVN total flux & Recovered VLA & $T_{\rm b}$  & Spatial Resolution \\
 & ($\mu$Jy\,beam$^{-1}$) & attenuation & ($\mu$Jy\,beam$^{-1}$) & ($\mu$Jy) & flux fraction & (K) & (pc$^2$) \\
\hline
SMG06 & 10.5 & 0.98 & $306\pm 11$ & $306\pm11$ & 1.24    & $5.0\times 10^5$  & $218\times197$ \\
SMG10 & 10.3 & 0.85 & $<37$       & \ldots     & $<0.13$ & $<3.6\times 10^4$ & $227\times205$ \\
SMG11 & 10.6 & 0.83 & $83\pm 13$  & $92\pm23$  & 0.38    & $1.4\times 10^5$  & $224\times203$ \\ \hline
SMG01 & 10.7 & 0.97 & $<33$       & \ldots     & $<0.45$ & $<8.5\times 10^4$ & $181\times175$ \\
SMG02 & 11.7 & 0.97 & $<36$       & \ldots     & $<0.55$ & $<5.2\times 10^4$ & $200\times188$ \\
SMG04 & 14.3 & 0.98 & $<44$       & \ldots     & $<0.40$ & $<5.5\times 10^4$ & $188\times184$ \\ \hline
\end{tabular}
\label{tab:evn}
\end{center}
\end{table*}

That starburst emission is dominant in our sample is not necessarily
unexpected, given the predictions of current models of SMG formation
and evolution \citep[e.g.,][]{baugh2005, swinbank2008,
  narayanan2009.smg, narayanan2009.co,narayanan2009.dog, Dave2010} and
observations \citep[e.g.][]{chapman2004, alexander2005, valiante2007,
  younger2008highres, younger2010highres, pope2008b, biggs2008,
  menendez2007, menendez2009, ivison2010b} which suggest that the
luminosity of SMGs is dominated by star formation. However, it is not
necessarily consistent with observations of the far-IR/radio
correlation in these systems. While several authors have claimed that
the local correlation holds at high redshift
\citep[e.g.][]{garrett2002, appleton2004, Ibar2008,
  younger2008.egsulirgs, murphy2009}, SMGs have been found to fall
systematically below the local relation.  This is usually taken to be
indicative of AGN activity \citep[e.g.][]{condon1992} and therefore
potentially in conflict with a diagnosis of star-formation-dominated
radio emission.

In Table~\ref{tab:fir.radio}, we summarise the far-IR/radio properties
for the sample presented in this work. All available far-IR photometry
was fitted with an isothermal greybody \citep{hildebrand1983} with
$S_\nu \sim \nu^\beta B_\nu(T_{\rm d})$ -- where $\beta$ is the dust
emissivity, $T_{\rm d}$ is the physical dust temperature, and $B_\nu$
is the Planck function -- and the errors in the fitted parameters were
estimated using a Monte Carlo simulation assuming Gaussian errors in
the photometric data.  This form of the greybody function assumes the
interstellar medium is optically thin in the far-IR, which may break
down in the most luminous starbursts \citep[e.g.,][]{scoville1991,
downes1998, papadopoulos10}, but while the fitted temperature is
sensitive to the the full emissivity term, $L_{\rm IR}$ and $q$ (the
logarithmic far-IR/radio ratio) are robust for $\nu_0 \gsim 1000$\,GHz
\citep[see also][]{yun2002}.

For SMG10 and SMG11, which have only one photometric point available,
we have fixed $T_{\rm d} = 35$\,K, appropriate for typical SMGs
\citep{kovacs2006, coppin2008.sharcii}, and for SMG11 we have assumed
$z=2.2^{+0.6}_{-0.5}$ \citep[the median and interquartile range for
SMGs with spectroscopic redshifts --][]{chapman2005}. We have also
fixed $\beta=1.5$, which is consistent with local ULIRGs
\citep{dunne2000, yang2007b} as well as current dust models
\citep{weingartner2001}. For the radio, we adopt radio spectral slopes
derived from 610-MHz maps obtained with the GMRT \citep{Ibar2009,
Ibar2010} in conjunction with the VLA 1.4-GHz data. As with the mean
$q$ estimated for SMGs with 350-$\mu$m photometry \citep[$q=2.09\pm
0.09$:][]{kovacs2006}, the objects in this sample fall below the local
value.  In fact, the mean for the star-formation-dominated SMGs with
no detection in EVN imaging is $\langle q \rangle=1.88\pm 0.22$; well
below the local value of $q=2.34$ \citep{yun2001}.


\begin{table*}
\begin{center}
\caption{Derived dust temperatures ($T_{\rm d}$), luminosities and
  far-IR/radio ratio ($q$) for each SMG. We define the far-IR
  luminosity between rest-frame 40 and 1000~$\mu$m. For SMGs 10 and
  11, only one photometric point was available in the far-IR and so we
  fixed $T_{\rm d}$ to 35\,K, consistent with the mean SMG far-IR
  spectral energy distribution inferred from 350--1200~$\mu$m
  photometry \citep{kovacs2006, coppin2008.sharcii}. As a
  spectroscopic redshift was not available for SMG11, we have assumed
  $z=2.2^{+0.6}_{-0.5}$ \citep[the median and interquartile range for
    radio-detected SMGs --][]{chapman2005} and propagate those
  uncertainties into the quoted errors on the fitted parameters. Also
  shown are the derived stellar masses, calculated following
  \citet{dye2008} and plotted in Fig.~\ref{fig:magorrian}.}
\begin{tabular}{cccccc} \hline
Name & $T_{\rm d}$ & $\log (L_{\rm IR}/\mathrm{L}_{\sun})$ & $q$ & $\log (M_{\mathrm{IR}}/\mathrm{M}_{\sun}$) \\
& (K) &  &  & \\
\hline
SMG06 & $56\pm 15$ & $13.1\pm 0.2$ & $2.32\pm 0.25$ & $11.6\pm0.2$ \\
SMG10 & $35$       & $12.7\pm 0.1$ & $1.93\pm 0.12$ & $11.7\pm0.2$ \\
SMG11 & $35$       & $12.8\pm0.1$  & $1.45\pm 0.35$ & $12.7\pm0.1$ \\ \hline
SMG01 & $28\pm 2$  & $12.8\pm 0.1$ & $1.99\pm 0.23$ & $12.2\pm0.7$ \\
SMG02 & $22\pm 3$  & $12.0\pm 0.2$ & $1.41\pm 0.33$ & $11.9\pm0.1$ \\
SMG04 & $29\pm 10$ & $12.0\pm 0.6$ & $1.80\pm 0.74$ & $11.53\pm0.02$ \\ \hline
\end{tabular}
\label{tab:fir.radio}
\end{center}
\end{table*}

There are three possible explanations for this discrepancy. First,
owing to a combination of sparse sampling and bad approximations
\citep[e.g.\ an isothermal dust population; see][]{Clements2010}, it
may be that the far-IR luminosity is systematically underestimated and
these objects do, in fact, fall on the local relation. Second, despite
some observational evidence to the contrary
\citep[e.g.,][]{garrett2002, appleton2004, younger2008.egsulirgs}, the
far-IR/radio correlation could evolve with redshift and/or environment
\citep{ivison2010a}. Finally, because SMGs are thought to be driven by
major mergers \citep{greve2005, tacconi2006, tacconi2008,
  younger2008highres, younger2010highres, narayanan2009.smg,
  narayanan2009.co}, there could be luminous synchrotron bridges
analogous to those seen in interacting systems in the local Universe
\citep{condon1993, condon2002, murphy2009}. We note that there are
also several more prosaic explanations, e.g.\ sample bias, flux
boosting and/or mis-identifications in the radio waveband.

\begin{figure}
\begin{center}
\includegraphics[scale=0.48]{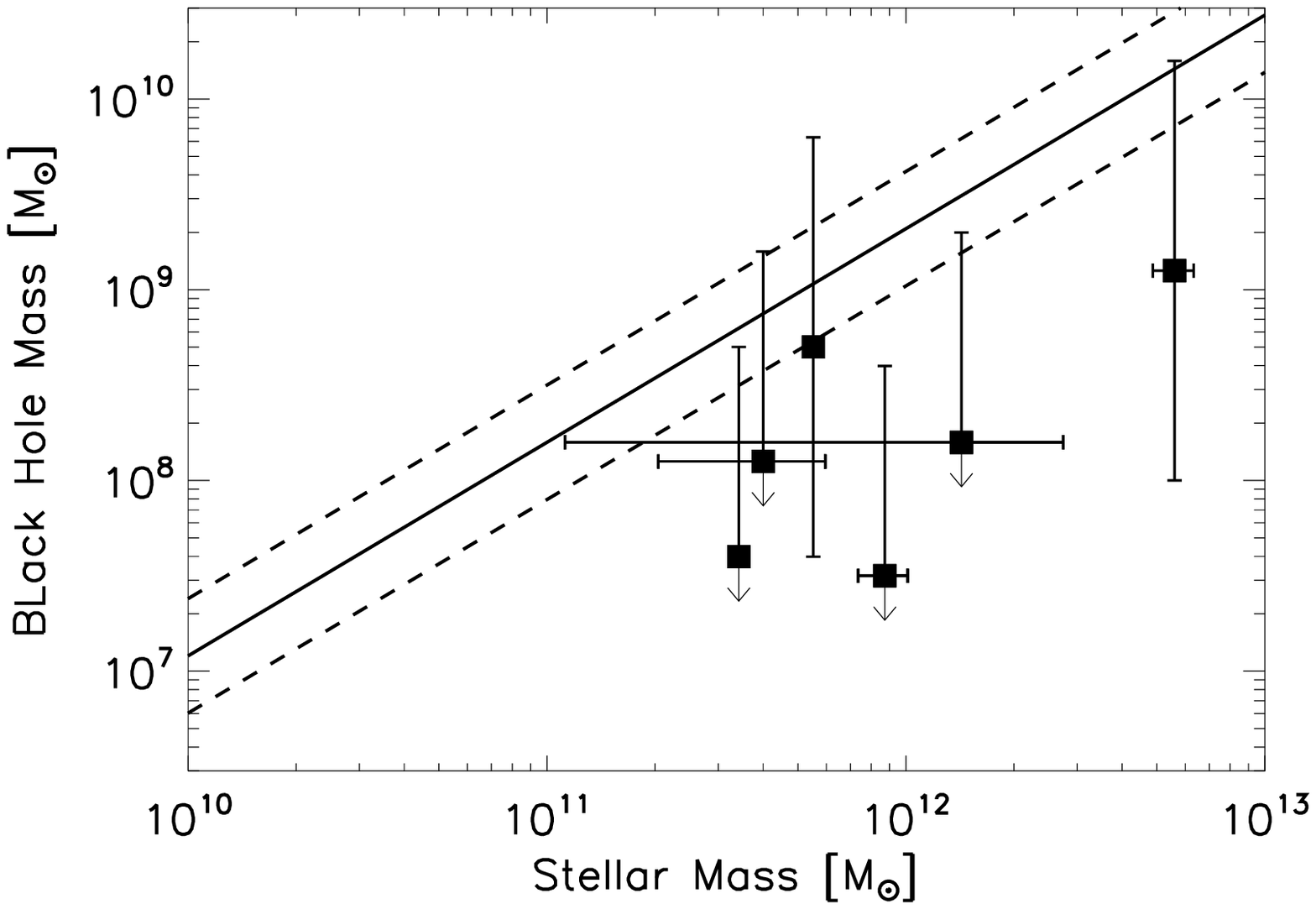}
\caption{SMBH mass versus stellar mass for all six SMGs, with black
  hole masses estimated from the 5-GHz radio luminosity
  \citep{lacy2001} and stellar masses estimated from fitting SED
  templates to the optical--IR photometry \citep{dye2008}.  The local
  relation from \citet{haring2004} is denoted by the solid line and
  its intrinsic scatter as dashed lines. Errors on the black hole
  masses are equal to the scatter in the \citet{lacy2001} relationship
  (1.1~dex) and those on the stellar masses are calculated as
  documented in \citet{dye2008}. At least one of the EVN non-detected
  SMGs is not consistent with the local relation and, due to being
  upper limits, all of the non-detections might signify black holes
  that are undermassive compared to the nearby universe.}
\label{fig:magorrian}
\end{center}
\end{figure}

For those targets with detections -- SMG06 and SMG11 -- we can use
their inferred radio luminosity to estimate the SMBH mass. Radio
observations of quasars -- including both radio-loud and radio-quiet
objects -- have established a strong correlation between their radio
luminosity and SMBH mass \citep{lacy2001}. Assuming that this relation
holds for AGN in heavily obscured environments, and that accretion is
Eddington-limited \citep[which is expected during the peak of both
  starburst and SMBH activity from merger-driven models --
][]{SDH2005}, then this observed relation is:
\[
\log(M_{\rm BH}/\mathrm{M}_{\sun}) = 0.52 \times {\rm log}(L_{\rm 5GHz}/\, {\rm W\, Hz^{-1}\,sr^{-1}})- 4.2
\]
where $L_{\rm 5GHz}$ is measured in the rest frame. Though the
intrinsic scatter is large -- approximately 1.1\,dex in $M_{\rm BH}$
at fixed $L_{\rm 5GHz}$ -- we can use these observations to provide a
rough estimate of the masses of SMBHs detected in the EVN imaging.
Again, for the radio we adopt the spectral slopes determined using
610-MHz maps obtained with the GMRT \citep{Ibar2009, Ibar2010} and
estimate black hole masses for SMG06 and SMG11 of ${\rm log}(M_{\rm
  BH}/\mathrm{M}_{\sun}) = $ 8.7 and 9.1 respectively. Errors on these are taken
to be equal to the intrinsic scatter (1.1\,dex) in the \citet{lacy2001}
relationship.

At the same time, for the non-detections (SMG10, SMG01, SMG02, and
SMG04), we can utilise the observed correlation to derive upper limits
on $M_{\mathrm{BH}}$ for these systems. This again requires we assume that no
radio-quiet AGN are present, and that emission from large-scale jets
is both negligible in these systems and in driving the observed
$M_{\mathrm{BH}}-L_{\rm 5GHz}$ correlation, and therefore represents a fairly
rough estimate. However, keeping these assumptions in mind we find
the 3--$\sigma$ detections for SMG10, SMG01, SMG02 and SMG04
imply upper limits of ${\rm log}(M_{\rm BH}/\mathrm{M}_{\sun}) \lsim$ 8.1,
8.2, 7.5 and 7.6, respectively.

Stellar masses and their associated uncertainties
(Table~\ref{tab:fir.radio}) were then estimated from model fits to the
observed optical--IR photometry following the methodology of
\citet{dye2008}. The resulting correlation with SMBH mass is presented
in Fig.~\ref{fig:magorrian} along with the empirical relationship
determined by \citet{haring2004} for a sample of 30 nearby
galaxies. While there is considerable uncertainty in the inferred SMBH
masses (and also in the assumed Eddington ratio) owing to the large
intrinsic scatter in the observed radio luminosity-black hole mass
relation, we find that the two SMGs with EVN detections are
statistically consistent with the local relation to within the
systematic uncertainty in the black hole mass estimates.

Of the non-detections, the predicted black hole masses are lower
than for the EVN-detected SMGs. Only SMG10 is clearly consistent with
the \citet{haring2004} local relationship although the uncertainties
are such that two others could also be classed as consistent. In only
one case (SMG02) does it appear likely that the black hole is
undermassive compared to the theoretical relation. However, as they are
upper limits, it cannot be ruled out that all are undermassive, a
result that would be consistent with the conclusions of
\citet{alexander2008} using X-ray photometry, and with the
expectations of current theoretical models \citep{narayanan2009.smg,
  narayanan2009.co} i.e.\ that submillimetre selection preferentially
identifies starbursts, prior to the peak of SMBH activity in massive,
gas-rich systems.

\section{Conclusions}
\label{sec:conclude}

We have undertaken a program designed to uncover AGN radio cores in a
sample of SMGs in the Lockman Hole, using very-long-baseline
interferometry with the EVN. Our sensitive, high-resolution images
have detected only two of our six targets, despite a strong selection
bias in favour of compact emission. From their brightness
temperatures, there can be little doubt that we are seeing radio
emission from active nuclei. For the other four SMGs we place upper
limits on the radio flux density of any compact component associated
with an AGN and conclude that star formation is probably the dominant
source of their radio emission.

\section*{Acknowledgements}

The European VLBI Network is a joint facility of European, Chinese,
South African and other radio astronomy institutes funded by their
national research councils. This work has benefited from research
funding from the European Community's sixth Framework Programme under
RadioNet R113CT 2003 5058187. This work has also been supported by the
European Community Framework Programme 7, Advanced Radio Astronomy in
Europe, grant agreement nl.: 227290. JDY acknowledges support from
NASA through Hubble Fellowship grant \#HF-51266.01, awarded by the
Space Telescope Science Institute, which is operated by the
Association of Universities for Research in Astronomy, Inc., for NASA,
under contract NAS 5-26555. RJI acknowledges an RCUK Individual Merit
award. We thank the referee for their helpful comments.

\bibliographystyle{mnras}
\bibliography{smg.new}

\end{document}